\documentclass[12pt]{article}

\usepackage{epsf}
\usepackage{epsfig}
\usepackage{cite}
\usepackage{amsmath}
\usepackage{graphics}

\begin{document}

\setlength{\textheight}{21.5cm}
\setlength{\oddsidemargin}{0.cm}
\setlength{\evensidemargin}{0.cm}
\setlength{\topmargin}{0.cm}
\setlength{\footskip}{1cm}
\setlength{\arraycolsep}{2pt}

\renewcommand{\thefootnote}{\#\arabic{footnote}}
\setcounter{footnote}{0}

\newcommand{\gtrsim}{ \mathop{}_{\textstyle \sim}^{\textstyle >} }
\newcommand{\lesssim}{ \mathop{}_{\textstyle \sim}^{\textstyle <} }
\newcommand{\rem}[1]{{\bf #1}}
\renewcommand{\thefootnote}{\fnsymbol{footnote}}
\setcounter{footnote}{0}
\def\thefootnote{\fnsymbol{footnote}}

\hfill {\tt arXiv:08mm.nnnn[hep-ph]}\\
\vskip .5in

\begin{center}

\bigskip
\bigskip

{\Large \bf Renormalizable $A_4$ Model for Lepton Sector}

\vskip .45in

{\bf Paul H. Frampton and Shinya Matsuzaki} 

\vskip .3in

{\it Department of Physics and Astronomy, University of North Carolina,
Chapel Hill, NC 27599-3255.}

\end{center}

\vskip .4in 
\begin{abstract}
We study flavor symmetry for the lepton sector with minimal Higgs
sector, namely one $A_4-$triplet SU(2)-doublet scalar.
To increase predictivity even further,
we impose the constraint of renormalizability.  
A geometric interpretation of an $A_4-$triplet aids our understanding
of tribimaximal mixing. 
We investigate the neutrino mass hierarchy
in such a minimal $A_4$ model and find there are two solutions:
one with $m_2 \gg m_1 = m_3$ is phenomenologically unacceptable; the
other with $m_3 \gg m_1 = m_2$ is a normal hierarchy. An inverted hierarchy
is impossible without addition of more parameters by
either more Higgs scalars or higher-order irrelevant operators.
\end{abstract}

\renewcommand{\thepage}{\arabic{page}}
\setcounter{page}{1}
\renewcommand{\thefootnote}{\#\arabic{footnote}}

\newpage

\noindent {\it Introduction}. ~~ 

\bigskip

In particle theory phenomenology, model building fashions
vary with time and because the present lack of data
(soon to be compensated by the Large Hadron Collider)
does not allow discrimination between models some fashions
develop a life of their own. In the present Letter
we take the apparently retrogressive step of imposing
the requirement of renormalizability\footnote{The model 
we discuss has an anomaly which can be cancelled in a more complete
$T^{'}$ model incorporating quarks without affecting
the results for leptons discussed in the present article.}
, as holds
for quantum electrodynamics (QED), quantum chromodynamics (QCD)
and the standard electroweak model, to show that
non-abelian flavor symmetry becomes then
much more restrictive and predictive. In a specific model
we show that a normal neutrino mass hierarchy is 
strongly favored over an inverted hierarchy.

\bigskip

For several years now there has been keen interest in the use
of $A_4$ \cite{Ma,A4} as a finite flavor symmetry in the lepton
sector, especially neutrino mixing; other approaches
are in \cite{Shelly}. In particular, the empirically approximate
tribimaximal mixing\cite{HPS} of the three neutrinos can 
be predicted. It is usually stated that either normal or 
inverted neutrino mass spectrum can be predicted.

\bigskip

In the present Letter we revisit these two questions in a
minimal $A_4$ framework with
only one $A_4$-{\bf 3} of Higgs doublets 
coupling to neutrinos and permitting
only renormalizable couplings. For such a minimal model
there is more predicitivity regarding neutrino masses.

\bigskip

Although the standard model was originally discovered
using the criterion of renormalizability, it is sometimes 
espoused\cite{Weinberg} that renormalizability
is not prerequisite in an effective lagrangian.
Nevertheless, imposing renormalizability in 
the present case is more sensible because it
does render the model far more predictive
by avoiding the many additional parameters
associated with higher-order irrelevant operators.
Our choice of Higgs sector also 
minimizes the number of free parameters.

\bigskip

We shall first discuss some geometry of $A_4$ symmetry as follows.

\bigskip
\bigskip

\noindent {\it Geometry of $A_4$ symmetry}

\bigskip

The group $A_4$ is the order g=12 symmetry of a
regular tetrahedron $T$ and is a subgroup of the
rotation group $SO(3)$. $A_4$ has irreducible representations
which are three singlets $1_1, 1_2, 1_3$ and a triplet $3$.
In the embedding $A_4 \subset SO(3)$ the {\bf 3} of $A_4$
is identified with the adjoint {\bf 3} of $SO(3)$.

\bigskip

Since the only Higgs doublets coupling to neutrinos
in our model are in a {\bf 3} of $A_4$, it is very useful
to understand geometrically the three components
of a {\bf 3}. 

\bigskip

A regular tetrahedron has four vertices, four faces and
six edges. Straight lines joining the midpoints of
opposite edges pass through the centroid and form
a set of three orthogonal axes. Regarding the regular
tetrahedron as the result of cutting off the four odd
corners from a cube, these axes are parallel to the
sides of the cube (see Fig. 1). With respect to the regular
tetrahedron, a vacuum expectation value (VEV)
of the {\bf 3} such as $<{\bf 3}> = v (1, 1, -2)$,
as will be used,clearly breaks SO(3) to U(1) and 
correspondingly $A_4$ to $Z_2$, since it requires
a rotation by $\pi$ about the 3-axis to restore
the tetrahedron.

\bigskip

At the same time, we can understand the appearance
of tribimaximal mixing\cite{Lee} with matrix\cite{HPS}

\begin{equation}
U_{TBM} = \left(
\begin{array}{ccc}
- \sqrt{\frac{1}{6}} & -\sqrt{\frac{1}{6}} & \sqrt{\frac{2}{3}} \\
\sqrt{\frac{1}{3}} & \sqrt{\frac{1}{3}} & \sqrt{\frac{1}{3}} \\
\sqrt{\frac{1}{2}} & - \sqrt{\frac{1}{2}} & 0 
\end{array}
\right),
\label{TBM}
\end{equation}
and our definitions are such that the ordering $\nu_{1, 2, 3}$
and $\nu_{\tau, \mu, e}$ satisfy
\begin{equation}
\left( \begin{array}{c} \nu_1 \\ \nu_2 \\ \nu_3 \end{array} \right)
= U_{TBM} 
\left( \begin{array}{c} \nu_{\tau} \\ \nu_{\mu} \\ \nu_e \end{array} \right)
\end{equation}
With respect to a side of the aforementioned cube,
a face diagonal is at angle $\pi/4$ and the body diagonal
is at angle $\tan^{-1} (\sqrt{\frac{1}{2}})$ with respect
to the body diagonal (see Fig. 2). These two angles, together with 
$\theta_{13} = 0$ are the three corresponding to 
the matrix of Eq.(\ref{TBM}). As we shall show this
mixing occurs naturally for VEVs $<{\bf 3}>
\propto (1, 1, 1)$  and $<{\bf 3}> \propto (1, 1, -2)$.

\bigskip

Assuming no CP violation, the Majorana matrix $M_{nu}$ is real and symmetric and
therefore of the form
\begin{equation}
M_{\nu} = \left(
\begin{array}{ccc}
A & B & C \\
B & D & F \\
C & F & E 
\end{array}
\right)
\label{ABCDEF}
\end{equation}
and is related to the diagonlized
form by
\begin{equation}
M_{diag} = \left( \begin{array}{ccc}
m_1 & 0 & 0 \\
0 & m_2 & 0 \\
0 & 0 & m_3
\end{array}
\right) = U_{TBM} M_{\nu} U_{TBM}^{T}.
\label{diag}
\end{equation}
Substituting Eq.(\ref{TBM}) into Eq.(\ref{diag}) shows
that $M_{\nu}$ must be of the general form in terms
of real parameters $A, B, C$:
\begin{equation}
M_{\nu} = \left(
\begin{array}{ccc}
A & ~~~ B & C \\
B & ~~~ A & C \\
C & ~~~ C & ~~~ A + B -C
\end{array}
\right),
\label{ABC}
\end{equation}
which has eigenvalues
\begin{eqnarray}
m_1 &=& (A + B - 2C) \nonumber  \\
m_2 &=& (A + B + C) \nonumber \\
m_3 &=& (A - B).
\label{masses}
\end{eqnarray}
The observed mass spectrum corresponds approximately
to $|m_1| = |m_2|$ which requires either $C=0$ or $C=2(A+B)$.
For a normal hierarchy, $(A+B)=0$ and $C=0$. For an inverted hierarchy
$A=B$ and $C=0$ or $C=4A$.

Now we study our minimal $A_4$ model to examine the occurrence
of the Majorana matrix Eq.(\ref{ABC}) and the eigenvalues
Eq.(\ref{masses}).

\bigskip

\noindent {\it Minimal $A_4$ model}

\bigskip

We assign the leptons to $(A_4, Z_2)$ irreps as follows
\footnote{These assignments differ from Ma-Rajasekharan \cite{Ma}
and from Altarelli-Feruglio \cite{A4}. 
The lagrangian, Eq. (\ref{lagrangian}) is, however, the most
general dimension-4 $(A_4 \times Z_2)$ invariant with our assignments.}

\begin{equation}
\begin{array}{ccc}
\left. \begin{array}{c}
\left( \begin{array}{c} \nu_{\tau} \\ \tau^- \end{array} \right)_{L} \\
\left( \begin{array}{c} \nu_{\mu} \\ \mu^- \end{array} \right)_{L} \\
\left( \begin{array}{c} \nu_e \\ e^- \end{array} \right)_{L}
\end{array} \right\}
L_L  (3, +1)  &
\begin{array}{c}
~~ \tau^-_{R}~ (1_1, -1)   \\
~~ \mu^-_{R} ~ (1_2, -1) \\
~~ e^-_{R} ~ (1_3, -1)  \end{array}
&
\begin{array}{c}
~~ N^{(1)}_{R} ~ (1_1, +1) \\
~~ N^{(2)}_R ~ (1_2, +1) \\
~~ N^{(3)}_{R} ~ (1_3, +1).\\  \end{array}
\end{array}
\end{equation}

\bigskip

The lagrangian is

\begin{eqnarray}
{\cal L}_Y
&=&
\frac{1}{2} M_1 N_R^{(1)} N_R^{(1)} + M_{23} N_R^{(2)} N_R^{(3)} \nonumber \\
& & + \Bigg\{
Y_{1} \left( L_L N_R^{(1)} H_3 \right) + Y_{2} \left(  L_L N_R^{(2)}  H_3
\right) + Y_{3} 
\left( L_L N_R^{(3)} H_3 \right)  \nonumber \\ 
&& +  
Y_\tau \left( L_L \tau_R H'_3 \right)
+ Y_\mu  \left( L_L \mu_R  H'_3 \right) + 
Y_e \left( L_L e_R H'_3 \right)  
\Bigg\}
+ 
{\rm h.c.} 
\label{lagrangian}
\end{eqnarray}
where $SU(2)$-doublet Higgs scalars are in $H_3(3, +1)$ and $H_3^{'}(3, -1)$.

\bigskip
\bigskip
\bigskip
\bigskip
\bigskip
\bigskip

The charged lepton masses originate
\footnote{The VEV of $H_3^{'}$ is chosen to give
charged leptons their masses and is not directly related 
to TBM mixing angles for neutrinos.}
from
$<H_3^{'}> = 
(\frac{m_{\tau}}{Y_{\tau}},\frac{m_{\mu}}{Y_{\mu}},\frac{m_{e}}{Y_{e}}) $ 
and are, to leading order, disconnected from the neutrino
masses if we choose a flavor basis where the charged leptons are mass eigenstates.
The $N_{R}^{i}$ masses break the $L_{\tau} \times L_{\mu} \times L_e$ symmetry but change the charged lepton masses
only by very small amounts $\propto  Y^2m_i/M_N$ at one-loop level.

\bigskip

The right-handed neutrinos have mass matrix
\begin{equation}
M_N =
\left(
\begin{array}{ccc}
M_1 & 0 & 0 \\
0 & 0 & M_{23} \\
0 & M_{23} & 0
\end{array}
\right).
\label{MN}
\end{equation}

\bigskip

We take the VEV of the scalar $H_3$ to be
\begin{equation}
<H_3> = (V_1, V_2, V_3),
\label{VEV}
\end{equation}
whereupon the Dirac matrix is
\footnote{The form of Eq.(\ref{MD}) follows from the
lagrangian of Eq.(\ref{lagrangian}) with general
$<H_3> = (V_1, V_2, V_3)$ and is not specialized to
either a Ma-Rajasekaran or an Altarelli-Feruglio basis.}

\begin{equation}
M_D = 
\left(
\begin{array}{ccc}
Y_{1} V_1 & ~~~ Y_{2} V_3 & ~~~ Y_3 V_2 \\
Y_1 V_3 & ~~~ Y_2 V_2 & ~~~ Y_3 V_1 \\
Y_1 V_2 & ~~~ Y_2 V_1 & ~~~ Y_3 V_3
\end{array}
\right).
\label{MD}
\end{equation}

\bigskip

\noindent The Majorana mass matrix $M_{\nu}$ is given by
\begin{equation}
M_{\nu} = M_D M_N^{-1} M_D^{T}.
\end{equation}
Defining $x_1 \equiv Y_1^2/M_1$ and $x_{23} \equiv Y_2Y_3/M_{23}$ we
find the symmetric form
\begin{equation}
\left(
\begin{array}{ccc}
x_1 V_1^2 + 2 x_{23} V_2V_3 & 
~~~ x_1V_1V_3 + x_{23} (V_2^2+V_1V_3) & 
~~~ x_1V_1V_2 + x_{23} (V_3^2+V_1V_2) \\
 & 
~~~ x_1V_3^2 + 2 x_{23} V_1V_2 & 
~~~ x_1V_2V_3 + x_{23} (V_1^2+V_2V_3)  \\ 
 & 
 & 
~~~ x_1V_2^2 + 2 x_{23} V_1V_3
\end{array}
\right).
\label{Mnu2}
\end{equation}

\bigskip

\noindent 
To ensure the texture of Eq.(\ref{Mnu2})
coincides with Eq.(\ref{ABC}), and hence gives tribimaximal
mixing we find the three equations corresponding
to the mixing angles

\begin{equation}
x_1V_1^2 + 2x_{23}V_2V_3 = x_1V_3^2 + 2x_{23}V_1V_2 
\label{A}
\end{equation}
\begin{equation}
x_1V_1V_2 + x_{23}(V_3^2+V_1V_2) = x_1V_2V_3 +x_{23}(V_1^2+V_2V_3)
\label{B}
\end{equation}
\begin{equation}
x_1(V_1^2+V_1V_3-V_1V_2) + x_{23}(2V_2V_3 +V_2^2 +V_1V_3 -V_3^2 - V_1V_2)
 = x_1V_2^2 + 2 x_{23} V_1V_3.
\label{C}
\end{equation}

\bigskip

We find no solutions of Eqs.(\ref{A},\ref{B},\ref{C}) with any of
$x_1, x_{23}, V_1, V_2, V_3$ vanishing. From 1-3 symmetry,
Eq. (\ref{A}) and Eq.(\ref{B}) are both satisfied only if
$V_1=V_3$. Solution of Eq.(\ref{C}) further requires
$(2V_1+V_2)(V_1-V_2) = 0$ since it can be shown that
$x_1=x_{23}$ is not possible for any hierachy
consistent with exeperiment.

\bigskip

In the minimal $A_4$ model, therefore, only two VEVs of
$H_3$ give tribimaximal mixing.

\bigskip

The first is
\footnote{This VEV, $<H_3> \propto (1, 1, 1)$, can be transformed
to $<H_3> \propto (0, 0, 1)$ by an $A_4$ transformation. These
possibilities have been called in the literature
the Ma-Rajaskaran and Altarelli-Feruglio bases respectively.
}
\begin{equation}
<H_3> = (V, V, V)
\label{VVV}
\end{equation}
studied in \cite{Lee}.
But careful comparison with the mass eigenstates reveals
that $m_2 \gg m_1=m_3=0$ so the wrong mass eigenstate ($\nu_2$)
is selected for the observed hierarchy, and so Eq.(\ref{VVV})
is an unacceptable VEV for $<H_3>$.  

\bigskip 

The only other VEV for the $A_4$-{\bf 3} is therefore
\footnote{Because $<H_3> \propto (1, 1, 1)$ could be made consistent
with the neutrino masses in most previous $A_4$ models
which had more parameters, the alternative 
$<H_3> \propto (1, -2, 1)$ seems not
to have been previously studied.}

\begin{equation}
<H_3> = (V, -2V, V),
\label{V2VV}
\end{equation}
which
also gives tribimaximal mixing and the mass spectrum $m_3 \gg m_1=m_2$
corresponding to a normal hierarchy. Thus Eq.(\ref{V2VV}) provides
the only allowed VEV for $<H_3>$.

\bigskip

An inverted hierarchy with $m_1=m_2 \gg m_3$ is not possible within
a minimal $A_4$ renormalizable model and so is disfavored. This
means, for example, that neutrinoless double $\beta$-decay
will require higher precision experiments.

\bigskip

Our conclusion is that 
the $A_4$ model in a minimal form does favor the normal hierarchy.
We have considered a more restrictive model based on $A_4$ than
previously considered\footnote{Previous analysis\cite{A4}
has included scalars called ``flavons" which are standard
model singlets with vacuum alignments
different from those of the present model in which 
scalars $H_3(Z_2=+1)$, $H_3^{'}(Z_2=-1)$ are electroweak
doublets.}. The theory has been required
to be renormalizable and the Higgs scalar
content is the minimum possible.

\bigskip

We have required that the neutrino mixing matrix be
of the tribimaximal form. We then find that the masses
for the neutrinos are highly constrained and
can be in a normal, not inverted hierarchy.

\bigskip

Most, if not all, previous $A_4$ models\cite{A4}
in the literature
permit higher-order irrelevant non-renormalizable
operators and their concomitant proliferation
of parameters and hence allow a wide variety if
possiblities
for the neutrino masses.
We believe the renormalizabilty condition is sensible
for these flavor symmetries because of the higher predictivity.

\bigskip

The next step, currently under intense investigation, is whether
the present renornmalizable $A_4$ model can be extended
to a renormalizable $T^{'}$ model\cite{FKTprime,Tprime}. It is necessary but
not
sufficent condition for this that a successful
renormalizable $A_4$ model, as presented here, exists.

\bigskip

Our principle conclusion is that the constraint of
renormalizability which is respected by successful
theories like QED and QCD gives sharper predictivity
to flavor symmetry. For example, in the model 
discussed in the present Letter, a normal
neutrino mass hierarchy is strongly
favored over an inverted neutrino mass hierarchy.
Such a conclusion cannot be reached without
invoking renormalizability as a working principle.

\bigskip
\bigskip
\bigskip
\bigskip

\newpage

\begin{center}

\section*{Acknowledgements}

\end{center}

This work was supported in part 
by the U.S. Department of Energy under Grant
No. DE-FG02-06ER41418.

\bigskip
\bigskip
\bigskip
\bigskip

\begin{center}

{\bf Figure captions}

\end{center}

\bigskip
\bigskip

\noindent Figure 1. Geometry of triplet of $A_4$.

\bigskip
\bigskip

\noindent Figure 2. Geometry of tribimaximal mixing.

\newpage

\begin{center} 

\includegraphics[scale=1.2]{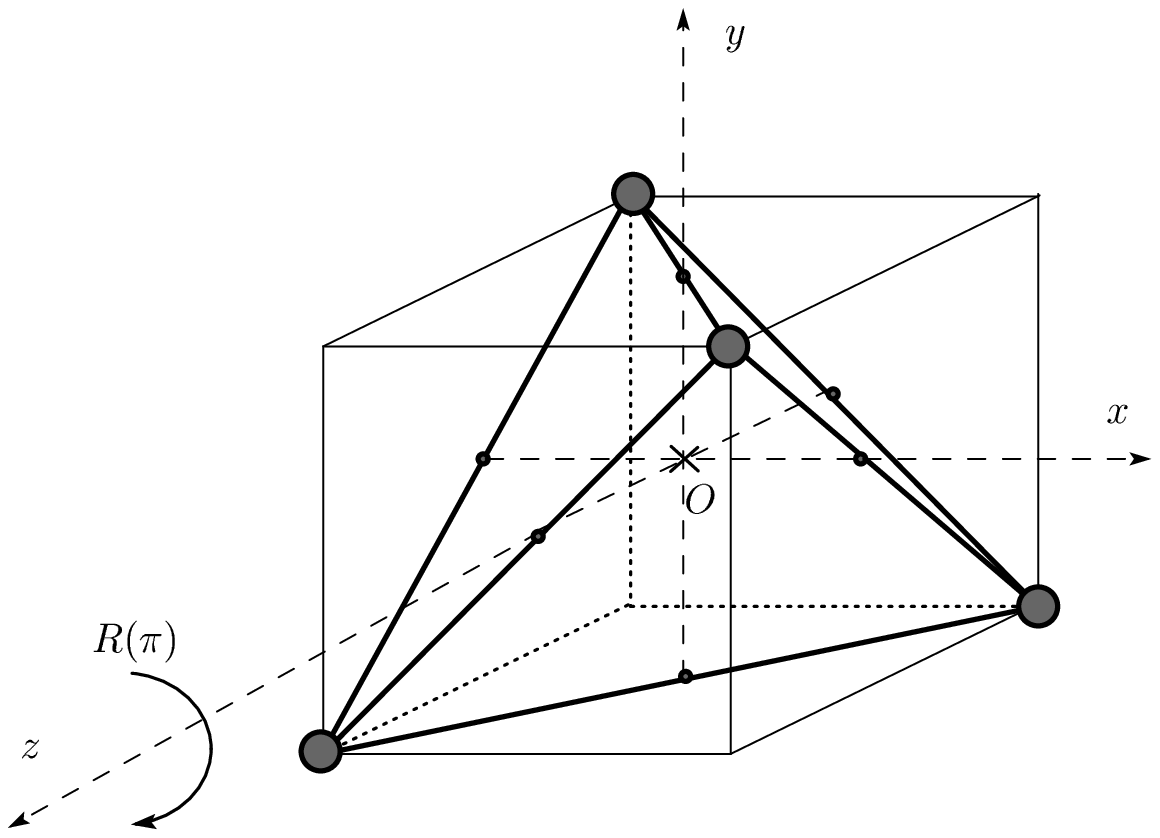}

\vspace{2.0in}

\noindent {\bf Figure 1.}

\end{center}

\newpage

\bigskip

\begin{center} 

\includegraphics[scale=1.5]{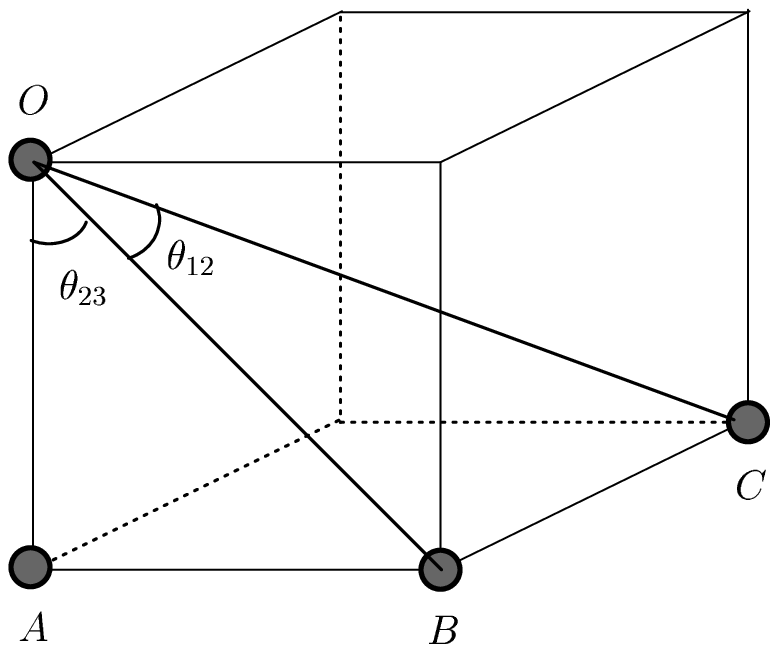}

\vspace{3.0in}

\noindent {\bf Figure 2.} 

\end{center}

\end{document}